\documentclass[11pt,twoside]{article}
\usepackage{graphicx,epsfig,natbib,epstopdf}
\usepackage{CS18}

\newcommand{\logg}{$\log g$}
\newcommand{\teff}{$T_{\rm eff}$}
\newcommand{\msun}{M$_\odot$}
\newcommand{\fl}{$F_8$}

\begin{document}
%
%
%
\title{Convection in Cool Stars, as Seen Through {\it Kepler}'s Eyes}
%
%
\author{Fabienne A. Bastien$^{1,}$$^{2}$}
\affil{$^1$Vanderbilt University, 1807 Station B, Nashville, TN 37235, USA}
\affil{$^2$Current Affiliation: Hubble Fellow, Pennsylvania State University, 525 Davey Lab, University Park, PA 16803, USA}
\begin{abstract}
Stellar surface processes represent a fundamental limit to the detection of extrasolar planets with the currently most heavily-used techniques.  As such, considerable effort has gone into trying to mitigate the impact of these processes on planet detection, with most studies focusing on magnetic spots.  Meanwhile, high-precision photometric planet surveys like CoRoT and {\it Kepler} have unveiled a wide variety of stellar variability at previously inaccessible levels.  We demonstrate that these 
newly revealed variations are not solely magnetically driven but also trace surface convection through light curve ``flicker.''  We show that ``flicker'' not only yields a simple measurement of surface gravity with a precision of $\sim$0.1 dex, but it may also improve our knowledge of planet properties, enhance radial velocity planet detection and discovery, and provide new insights into stellar evolution.
\end{abstract}
%
%
%
%
%
\section{Introduction}

Most planets are observed only indirectly, through their influence on their host star.  The planet properties we infer therefore strongly depend on how well we know those of the stars.  Our ability to determine the surface gravity (\logg) of field stars, however, is notoriously limited: broadband photometry, while efficient, yields errors of $\sim$0.5 dex; spectroscopy suffers from well known degeneracies between \logg, \teff and metallicity \citep{torres10} while having \logg\ errors of 0.1--0.2 dex \citep{ghezzi10}; and asteroseismology, the gold standard for stellar parameter estimation with \logg\ errors of $\sim$0.01 dex \citep{chaplin11,chaplin14}, is time and resource intensive and, particularly for dwarfs, is limited to the brightest stars.

Meanwhile, high precision photometric surveys like CoRoT and {\it Kepler} have surveyed over $\sim$200 000 Sun-like stars in their hunt for exoplanets, revealing stellar variations that have previously only been robustly observed in the Sun and a handful of bright Sun-like stars --- and also variations that were previously unknown but, as we show, encode a simple measure of stellar \logg.  In what follows, I describe our analysis of the newly unveiled high frequency photometric variations, which we term ``flicker'' (or \fl) and which enable us to measure \logg\ with an accuracy of $\sim$0.1~dex.  I summarize our work thus far in using \fl\ to study granulation in Sun-like stars, to examine the impact of granulation in radial velocity planet detection, and to improve size estimates of transiting exoplanets.

\section{Photometric ``Flicker:'' a Tracer of Granulation and a Simple Measure of Stellar Surface Gravity}

\begin{figure}
\centering
\includegraphics[scale=.30]{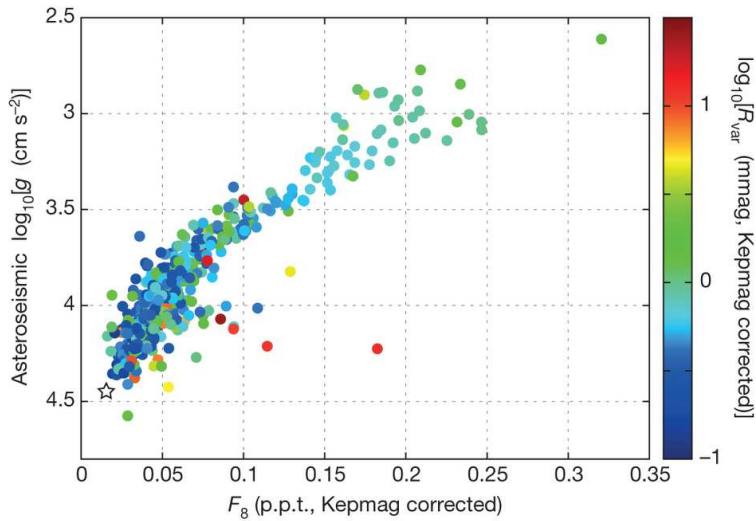}
\caption{\label{fig:bastien13}
Stellar surface gravity manifests in a simple measure of brightness variations. Asteroseismically determined \logg\ shows a tight correlation with \fl. Color represents the amplitude of the stars' brightness variations; outliers tend to have large brightness variations. Excluding these outliers, a cubic-polynomial fit through the Kepler stars and through the Sun (large star symbol) shows a median absolute deviation of 0.06~dex and a r.m.s.\ deviation of 0.10~dex. To simulate how the solar \logg\ would appear in data we use to measure \logg\ for other stars, we divide the solar data into 90-d ``quarters''. Our \fl--\logg\ relation measured over multiple quarters then yields a median solar \logg\ of 4.442 with a median absolute deviation of 0.005~dex and a r.m.s.\ error of 0.009~dex (the true solar \logg\ is 4.438). From \citet{bastien13}.}
\end{figure}

Using light curves from NASA's {\it Kepler} mission, we discovered that stellar \logg\ reveals itself through \fl – a measure of photometric variations on timescales of $<$ 8hr --- and may hence be used to measure \logg\ with errors of $\sim$0.1 dex, even for stars too faint for asteroseismology \citep[Fig.~\ref{fig:bastien13}]{bastien13}.  The measurement of \logg\ from \fl\ only requires the discovery light curves, and this measurement not only yields a result with an accuracy that rivals spectroscopy, it also does so very quickly and efficiently, requiring only a simple routine that can be executed by anyone in just a few seconds per star.

In \citet{bastien13}, we ascribed \fl\ to granulation power, which is known to depend on the stellar \logg\ \citep{kjeldsen11,mathur11}.  Recent independent simulations and asteroseismic studies have examined the expected photometric manifestations of granulation \citep{samadi13a,samadi13b,mathur11}, nominally through the Fourier spectrum from which it can be difficult to extract the granulation signal.  We used the simulations to predict the granulation-driven \fl, and we find excellent agreement with our observed \fl, demonstrating that the \fl\ is indeed granulation-driven \citep{cranmer14}. We also determined an empirical correction to the granulation models, particularly for F stars which have the shallowest convective outer layers. Indeed, our results suggest that these models must include the effects of the magnetic suppression of convection in F stars in order to reproduce the observations.  This work can ultimately help to develop our technique of “granulation asteroseismology,” enabling the precise determination of a larger number of stellar, and hence planetary, parameters.

\section{Stellar ``Flicker'' Suggests Larger Radii for Bright {\it Kepler} Planet Host Stars}

\begin{figure}
\centering
\includegraphics[scale=.30]{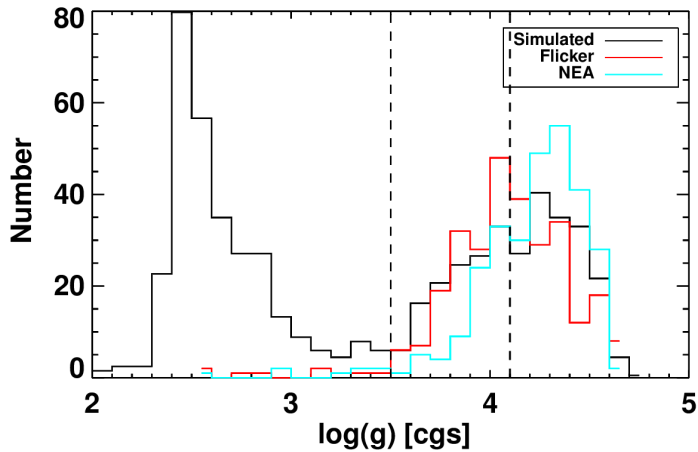}
\caption{\label{fig:bastien14b_1}
Distributions of \logg\ for the TRILEGAL simulated sample (black) and KOI host stars with \fl-based \logg\ (red) and broadband photometry/spectroscopy-based \logg\ (``NEA''; cyan curve). We limit the \teff\ range here to 4700--6500~K, for which the {\it Kepler} targets should be representative of the field. Vertical lines indicate the range of \logg\ corresponding to subgiants.  We find that \fl\ reproduces the expected underlying distribution, and, in particular, recovers the expected population of subgiants, while the NEA parameters are preferentially pushed towards the main sequence.  Adapted from \citet{bastien14b}.}
\end{figure}

\begin{figure}
\centering
\includegraphics[scale=.50]{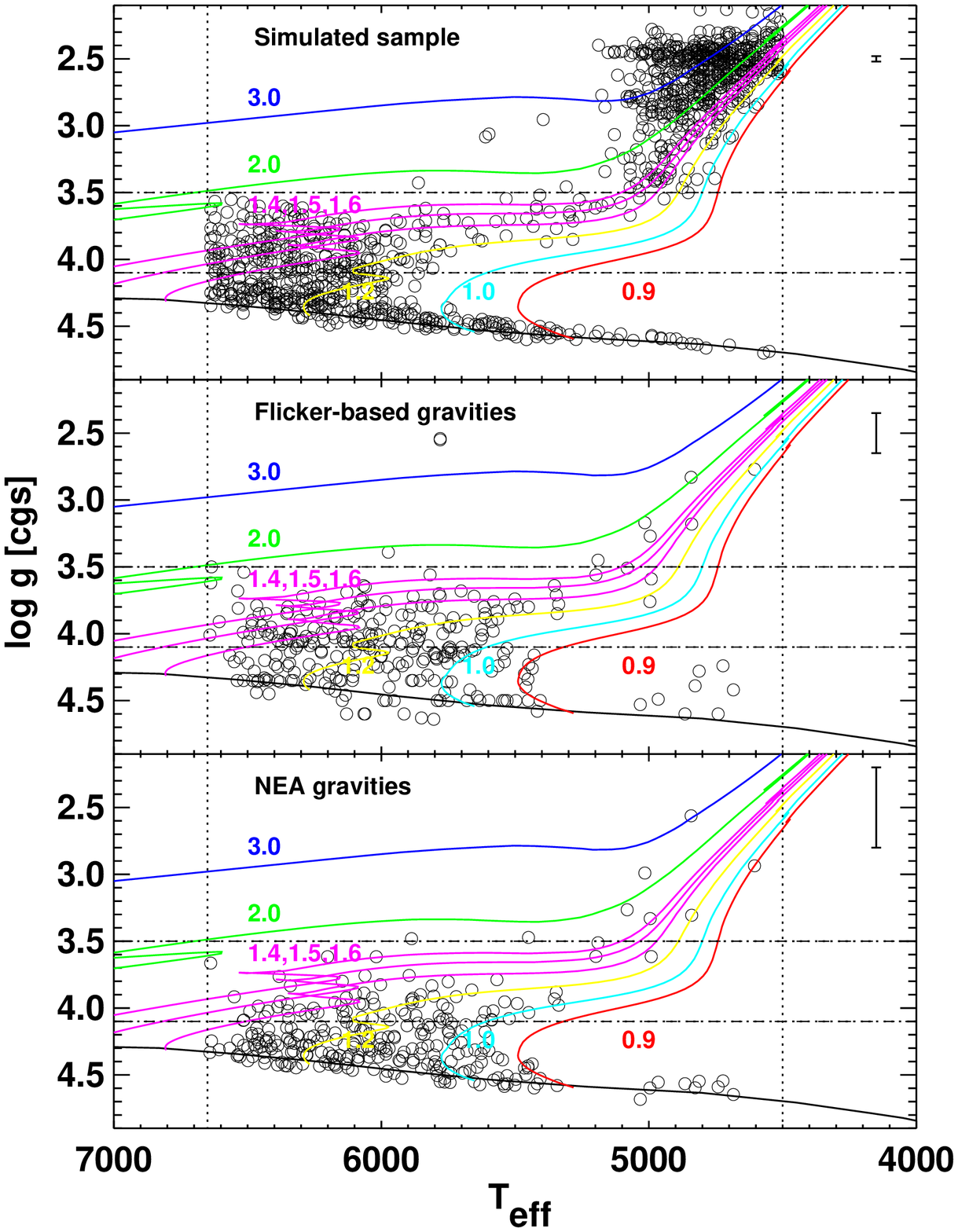}
\caption{\label{fig:bastien14b_2}
H-R diagram of KOI host stars with \logg\ derived from \fl\ (middle) and broadband photometry/spectroscopy (bottom), and as predicted by a TRILEGAL \citep{girardi05} simulation (top). Colored curves represent the theoretical evolutionary tracks (masses labeled in \msun). Vertical lines demarcate the range of stellar \teff considered in this study. The horizontal lines demarcate the range of \logg\ for subgiants (3.5 $<$ \logg\ $<$ 4.1). A representative error bar on \logg\ for each stellar sample is in the upper right of each panel. We find that the \fl-based \logg\ distribution more closely matches expectation than previous \logg\ measurements, particularly in the subgiant domain, perhaps because \fl involves no main-sequence prior on the \fl-based \logg\ values.  From \citet{bastien14b}.}
\end{figure}

The speed and efficiency with which one can determine accurate \logg\ solely with the discovery light curves translates directly into a rapid assessment of the distribution of bulk planet properties --- in particular, with greater accuracy and fewer telescopic and computational resources than similar studies \citep{batalha13,burke14} that of necessity relied on broadband photometric measurements to determine stellar properties.  We therefore applied our \fl\ technique to a few hundred bright ({\it Kepler} magnitudes between 8 and 13) planet candidates in the {\it Kepler} field, and we find that these stars are significantly more evolved than previous studies suggest \citep{bastien14b}.  As a result, the planet radii are 20--30\% larger than previously estimated.  In addition, we find that the high proportion of subgiants we derive (48\%) is consistent with predictions from galactic models of the underlying stellar population (45\%), whereas previous analyses heavily bias stellar parameters towards the main sequence and hence yield a low subgiant fraction (27\%; Figs.~\ref{fig:bastien14b_1},\ref{fig:bastien14b_2}).

     We expand upon this work by tailoring our initial \fl\ relation to be more directly useful to the exoplanet community by deriving a relationship between \fl\ and stellar density \citep{kipping14}.  This relation, which can yield the stellar density with an uncertainty of $\sim$30\%, can help to constrain exoplanet eccentricities and enable the application of techniques like astrodensity profiling to hundreds of exoplanet host stars in the {\it Kepler} field alone.

\section{RV Jitter in Magnetically Inactive Stars is Linked to High Frequency ``Flicker'' in Light Curves}

RV planet detection, particularly of small planets, requires precise Doppler measurements, and only a few instruments are able to achieve the precision needed to observe them.  Key to the success of RV planet campaigns is the avoidance of ``RV loud'' stars --- those likely to exhibit large levels of RV jitter that can impede and sometimes even mimic planetary signals \citep{queloz01}.  Most RV surveys therefore focus their attention on magnetically quiet stars, as magnetic spots tend to drive the largest amount of RV jitter.  Nonetheless, magnetically inactive stars can exhibit unexpectedly high levels of RV jitter \citep{wright05,galland05}, and even low jitter levels can impede the detection of small planets.  The drivers of RV jitter in inactive stars remain elusive \citep{dumusque11a,dumusque11b,boisse12}, continuing to plague RV planet detection and, in the case of F dwarfs, resulting in the outright avoidance of whole groups of notoriously RV noisy stars, even in transit surveys with large ground-based follow-up efforts like {\it Kepler} \citep{brown11}.

\begin{figure}
\centering
\includegraphics[scale=.40, angle=90]{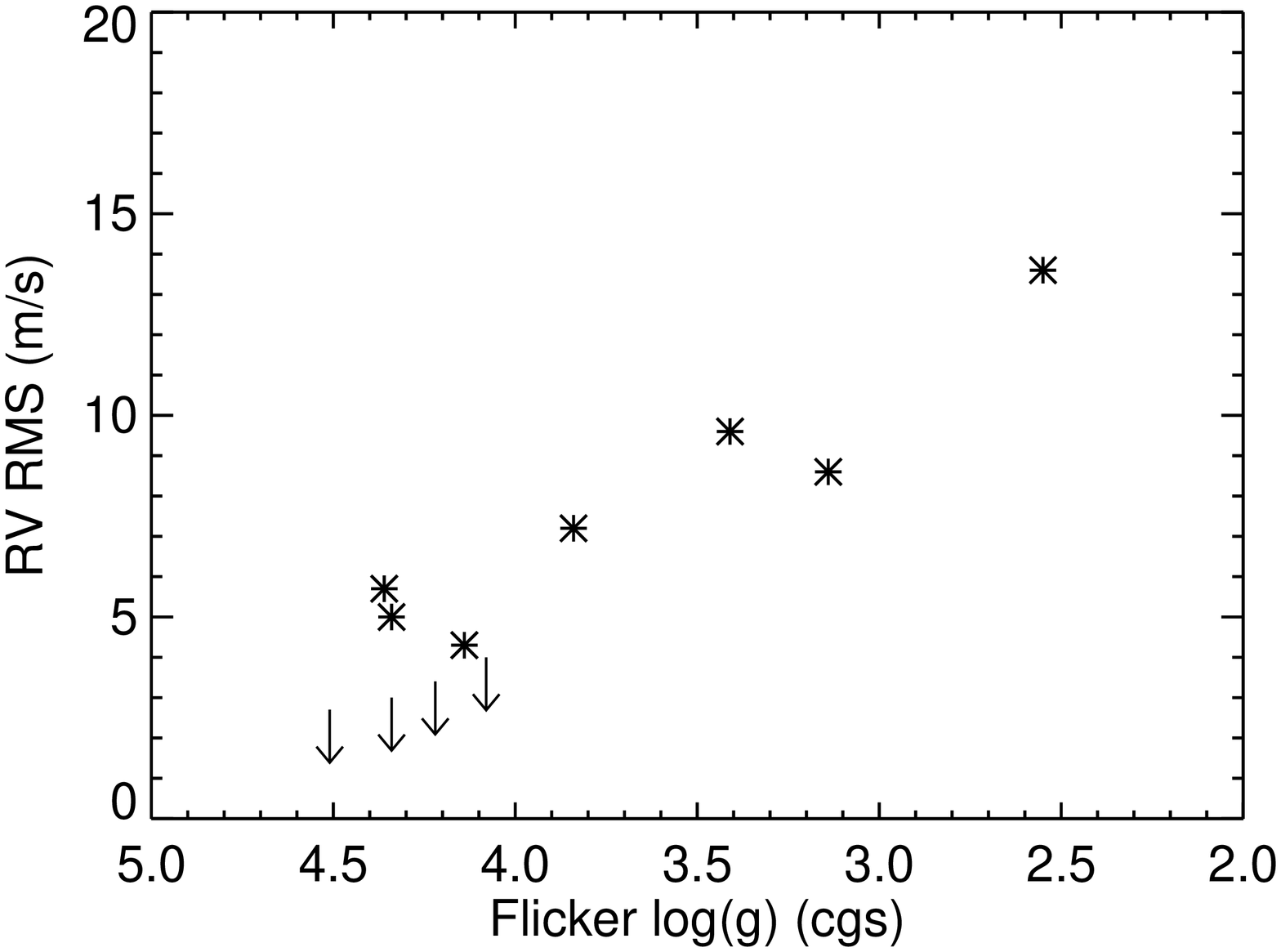}
\caption{\label{fig:bastien14a}
Comparison between RV jitter (RV RMS) and \fl-based \logg: RV jitter shows a strong anti-correlation with \fl-based \logg, with a statistical confidence of 97\% derived from a survival analysis. A similar trend was found by \citet{wright05}. \fl\ measures granulation power \citep{bastien13}, indicating that the RV jitter of magnetically inactive stars is driven by convective motions on the stellar surface whose strength increases as stars evolve.  Adapted from \citet{bastien14a}.}
\end{figure}

Given the breadth of stellar photometric behavior newly revealed by ultra-high precision light curves, and the new insights that they are giving into stellar surface processes, we compared different ways of characterizing this photometric behavior with RV jitter for all stars with both ultra-high precision light curves and high precision, long term RV monitoring \citep{bastien14a}.  These stars have very low photometric amplitudes (less than 3 ppt), a previously unexplored regime of both photometric variability and RV jitter.  We find that the RV jitter of these stars, ranging from 3~m~s$^{-1}$ to 135~m~s$^{-1}$, manifests in the light curve Fourier spectrum, which we then use to develop an empirical predictor of RV jitter.   We also find that spot models grossly under-predict the observed jitter by factors of 2--1000.    Finally, we demonstrate that \fl\ itself is a remarkably clean predictor of RV jitter in magnetically quiet stars (Fig.~\ref{fig:bastien14a}), suggesting that the observed jitter is driven by convective motions on the stellar surface and is strongly tied to \logg.

\section{Summary}

We find that surface convection in cool stars manifests as the high frequency ``flicker'' observed in high precision, long time-baseline light curves, such as those from {\it Kepler}.  We show that it yields a simple measure of stellar surface gravity and density, and we use it to place empirical constraints on granulation models.  We use it to perform an ensemble analysis of exoplanet host stars, finding that the exoplanet radii are larger than previous studies suggested.  Finally, we find that it is a clean predictor of RV jitter in magnetically inactive stars and can hence be used to identify promising targets for RV follow-up campaigns and RV planet searches.

More generally, we show that stellar variability --- traditionally considered a major noise source and nuisance, particularly in exoplanet detection --- can be used to enhance both exoplanet science and our understanding of stellar evolution.

\acknowledgments{
I thank the Cool Stars 18 SOC for a great conference and for kindly inviting me to share my work.  I would like to acknowledge funding support from a NASA Harriet Jenkins Pre-Doctoral Fellowship and helpful discussions with all those who have contributed to the results discussed here.
}

\end{document}